\documentclass[aps,prd,reprint,amsmath,amssymb,showpacs,superscriptaddress]{revtex4-1}
\usepackage{graphicx}
\usepackage{subfigure}
\usepackage{dcolumn}
\usepackage{bm}
\usepackage{slashed}
\usepackage{ulem}
\usepackage{epstopdf}
\usepackage{mathrsfs}
\usepackage{color,xcolor}

\usepackage{hyperref}

\begin{document}


\title{Heavy-light mesons beyond ladder approximation}

\author{Pianpian Qin} %
\affiliation{Department of Physics and State Key Laboratory of Nuclear Physics and Technology,\\
 Peking University, Beijing, 100871, China}

\author{Si-xue Qin} %
\email{sqin@cqu.edu.cn}
\affiliation{Department of Physics, Chongqing University, Chongqing 401331, P. R. China.}

\author{Yu-xin Liu} %
\email{yxliu@pku.edu.cn}
\affiliation{Department of Physics and State Key Laboratory of Nuclear Physics and Technology,\\
 Peking University, Beijing, 100871, China}
\affiliation{Collaborative Innovation Center of Quantum Matter, Beijing 100871, China}
\affiliation{Center for High Energy Physics, Peking University, Beijing 100871, China}

\date{\today}

\begin{abstract}
The heavy-light mesons are studied within the framework of Dyson-Schwinger equations of QCD. Inspired by the axial-vector Ward-Takahashi identity resulting from the chiral symmetry, we propose a truncation scheme beyond the ladder approximation without introducing any additional parameter.
For the pseudoscalar and vector heavy-light mesons, the obtained mass spectrum has the level of relative errors at $5\%$ compared with experimental data and lattice-QCD results.
For the leptonic decay constants, our results are comparable with those from experiments and/or lattice QCD. For some channels, the discrepancies are sizable but significantly smaller than those using the equal spacing rule.
The truncation scheme proposed in this work is simple and could be improved and applied to study other open flavor hadrons including both mesons and baryons.
\end{abstract}



\maketitle


\section{Introduction}\label{SecI}
In recent decades, many heavy-light mesons are observed in experiments~\cite{PDG2018}, which has attracted a lot of interest.
The study of heavy-light mesons provides a way to understand dynamical chiral symmetry broken (DCSB)~\cite{Eichten1974PRD,Pennington2005JPCS,Bashir2012PRC}--one of the most fascinating features of the quantum-chromo dynamics (QCD).
Nevertheless it is still very challenging.
The heavy-light mesons are two-body bound states of heavy and light (anti-)quarks, e.g., $c\bar{d}$, $c\bar{u}$, $c\bar{s}$, etc, which are highly flavor unsymmetric. Thus, the successful descriptions require a systematic and self-consistent understanding of strong interaction in both ultraviolet and infrared regions. Herein, non-perturbative approach of QCD is the key for solving the problem.

Several non-perturbative approaches of QCD have been developed, such as the lattice QCD (lQCD)~\cite{Wilson1974PRD,Ginsparg1982PRD,Batrouni1985PRD,Ryan2018LNP}, the Dyson-Schwinger equations (DSEs)~\cite{Williams1994PPNP,Roberts2000PPNP,Alkofer2001PR,Maris2003IJMP,Roberts2008PPNP,Bashir2012CTP,Cloet2014PPNP}, the functional renormalization group (FRG)~\cite{Polonyi2003CEJP,Pawlowski2007AP,Gies2012LNP}, and so on.
In this work, the QCD's DSE approach is implemented. Within the DSE framework, mesons are described by the two-body Bethe-Salpeter equation (BSE)~\cite{Salpeter1951PR,Nakanishi1969PTPS,Munczek1995PRD,Bender1996PLB,Alkofer2001PR,Qin2016FBS}.
As input, the quark propagator must be solved by the one-body gap equation.
The key point is that the one-body and the two-body equations must be constructed in a self-consistent way to preserve QCD's symmetries. For instance, the chiral symmetry is crucial to realize pion's twofold role as quark-antiquark bound state and Nambu-Goldstone boson~\cite{Munczek1995PRD,Bender1996PLB,Maris1998PLB}.

The simplest symmetry-preserving scheme is the so called rainbow-ladder (RL) truncation~\cite{Williams1994PPNP,Maris1997PRC,Maris1999PRC,Bhagwat2004PRC,Eichmann2008PRC,Qin2012PRC85.035202}, which approximates the dressed quark-gluon vertex as the bare one and the quark-antiquark scattering kernel as the one-gluon exchange form.
Since the bare vertex is flavor-blind and the one-gluon exchange lacks of spin-orbit repulsion, the RL truncation can work for ground state pseudoscalar and vector mesons with symmetric flavors,
e.g., the light-light mesons~\cite{Maris1997PRC,Maris1998PLB,Maris1999PRC,Qin2012PRC85.035202}
and the heavy-heavy mesons~\cite{Casalbuoni1997PR,Bali2001PR,Fischer2015EPJA,Ding2016EPJWC,Chen2017PRD}.

For heavy-light mesons, a sophisticated improvement of the RL truncation must be developed. Recently, following this direction, some progress has been made~\cite{Ivanov1999PRD,Rojas2014PRD,Gomez2015FBS,Gomez2015PRD,Gomez2016PRD,Hilger2017EPJA,Bedolla2018EPJC,Chen2019CPC,Binosi2019PLB,Yin2019PRD100.034008}. In this work, inspired by the axial-vector Ward Takahashi identity (AV-WTI)~\cite{Maris1998PLB,Maris1997PRC}, we propose a flavor-sensitive Bethe-Salpeter kernel beyond the simplest ladder approximation.
This kernel degenerates to the ladder approximation for flavor-symmetric systems, and thus can be considered as a generalization of the latter.
Without introducing any new parameter, the mass spectra and leptonic decay constants of heavy-light mesons in pseudoscalar and vector channels can be well reproduced, self-consistently and systematically.

The article is organized as follows:
In Sec.~\ref{SecII}, we describe the DSE approach and the setups.
In Sec.~\ref{SecIII}, we fit the parameters with the light-light and heavy-heavy systems first, and then present results of heavy-light mesons. The comparisons with the experimental data and results of lattice QCD are also included.
Section~\ref{SecIV} provides a summary and perspective.

\section{Theoretical Framework}\label{SecII}
\subsection{The Gap Equation}\label{SecIIA}

The quark gap equation reads~\cite{Williams1994PPNP}
\begin{equation}\label{DSE}
S^{-1}(p)=Z_2(i\slashed{p}+Z_m m)+\Sigma(p),
\end{equation}
with the self energy
\begin{equation}
\Sigma(p)= g^2 Z_1\int_q^{\Lambda} D_{\mu\nu}(p-q)  \frac{\lambda^a}{2}\gamma_{\mu}S(q) \frac{\lambda^a}{2}\Gamma_{\nu}(q,p),
\end{equation}
where $Z_1$, $Z_2$, and $Z_m$ are the vertex, quark wave-function, and mass renormalization constants, respectively; $m$ is the renormalized current quark mass; ${\lambda^a}$ are the color matrices; $\int_q^{\Lambda}$ represents a Poinc$\acute{a}$re invariant regularization of the four-dimensional integral, with $\Lambda$ the ultraviolet regularization mass scale; $\Gamma_{\nu}$ and $D_{\mu\nu}$ are the dressed quark-gluon vertex and the dressed gluon propagator, respectively.

The solution of the gap equation, i.e., the dressed quark propagator, can be decomposed as
\begin{eqnarray}
S(p)&=&-i \slashed{p} \sigma_v(p^2) + \sigma_s(p^2), \label{quark_lorentz1}\\
S^{-1}(p)&=&i \slashed{p} A(p^2) +B(p^2),    \label{quark_lorentz2}
\end{eqnarray}
with the momentum subtraction renormalization condition
\begin{equation}
S^{-1}(p)|_{p^2=\zeta^2}=i\slashed{\zeta} + m^\zeta,
\end{equation}
where $\zeta$ is the renormalization point and $m^\zeta$ the renormalized current-quark mass. The dressed quark mass function reads then
\begin{equation}\label{MassFunc}
 M(p^2)=B(p^2,\zeta^2)/A(p^2,\zeta^2),
\end{equation}
which is independent of the renormalization point $\zeta$. Note that $m^\zeta$ is the mass function evaluated at the renormalization point $m^\zeta=M(\zeta^2)$.

To solve the gap equation, the dressed quark-gluon vertex and the dressed gluon propagator must be specified. The rainbow part of the RL approximation can be expressed as $(k=p-q)$
\begin{align}
Z_1 g^2 D_{\mu\nu}(k)\Gamma_{\nu}(q,p) = k^2 \mathcal{G}(k^2)D_{\mu\nu}^{\text{free}}(k)\gamma_{\nu},\label{InteractionModel}
\end{align}
where $D_{\mu\nu}^{\text{free}}(k)=(\delta_{\mu\nu}-\frac{k_{\mu}k_{\nu}}{k^2})\frac{1}{k^2}$ is the Landau-gauge free gluon propagator. The interaction model $\mathcal{G}(k^2)$ is written as
\begin{align}
k^2 \mathcal{G}(k^2) = k^2\mathcal{G}_{IR}(k^2)+4\pi\tilde{\alpha}_{pQCD}(k^2)\,,
\end{align}
where $\tilde{\alpha}_{pQCD}(k^2)$ is a bounded and monotonically decreasing regular continuation of the perturbative QCD running coupling to all values of spacelike-$k^2$; and $\mathcal{G}_{IR}(k^2)$ is an $ansatz$ for the interaction at infrared region and dominates in the region $|k|<\Lambda_{QCD}$. The form of $\mathcal{G}_{IR}(k^2)$ determines whether the DCSB and/or confinement can be realized.

With the Qin-Chang (QC) model~\cite{Qin2011PRC}, the interaction is expressed as ($s=k^2$)
\begin{equation}\label{QC}
\mathcal{G}(s)=\frac{8\pi^2}{\omega^4}D e^{-s/\omega^2}+\frac{8\pi^2\gamma_m\mathcal{F}(s)}{\ln[\tau+(1+s/\Lambda_{QCD}^2)^2]},
\end{equation}
where $\mathcal{F}(s)=(1-e^{-s/4m_t^2})/s$ with $m_t=0.5$ GeV; $\gamma_m=12/(33-N_f)$ is the dimension anomaly with the flavor number $N_{f}$; $\tau=e^{2} -1$ is a constant.
Following Ref.~\cite{Qin2018PRD}, we take $N_{f} =5$ and $\Lambda_{QCD}=0.36\,$GeV.

The pointwise behavior of Eq.~\eqref{QC} can also be parameterized as~\cite{Aguilar2009PRD,Aguilar2012PRD}
\begin{equation}
\mathcal{G}(k^2)=\frac{4\pi\alpha(k^2)}{k^2+m_g^2(k^2)},\quad\quad m_g^2(k^2)=\frac{M_g^4}{k^2+M_g^2},
\end{equation}
where $m_g(k^2)$ is the gluon mass and $M_g$ is the mass scale. The interaction achieves its maximum value at the deep infrared $k^2=0$. The running behavior and the gluon mass scale are consistent with the modern DSE and lattice-QCD results \cite{Bowman2004PRD,Oliveira2011JPG,Binosi2015PLB,Binosi2017PRD,Rodriguez2018FBS}.

Many studies (e.g., Refs.~\cite{Qin2012PRC85.035202,Xin2014PRD,Gao2014PRD89.076009,Qin2016FBS,Chen2017PRD,Chen2018PRD,Qin2018PRD,Qin2019FBS}) have shown that the above mentioned interaction model is quite successful.
The parameters $D$ and $\omega$ control the strength and the width of the interaction, respectively.
In fact, observable properties of light-quark ground-state vector- and isospin-nonzero pseudoscalar mesons are insensitive to variations of $\omega \in[0.4,0.6]\,$GeV, as long as
\begin{equation}
\varsigma^3:=D\omega = {\rm constant}\,.
\end{equation}
Since the interaction model absorbs the dressing effects of the quark-gluon vertex and the gluon propagator, for light quark systems, it has much larger strength compared with that of the realistic gauge-sector interaction at infrared momenta~\cite{Qin2011PRC} and the one determined by solving the coulped equations of the quark propagator and the quark-gluon interaction vertex~\cite{Tang2019PRD}.
However, straightforward analysis shows that corrections to the RL truncation almost vanish in the heavy-quark limit~\cite{Bhagwat2004PRC}; hence, the aforementioned agreement entails that the RL truncation should provide a reasonable approximation for the systems involving only heavy quarks so long as one employs a smaller strength. This will be discussed in detail in Sec.~\ref{SecIII}.

\subsection{The Bethe-Salpeter Equation}\label{SecIIB}
In terms of Green functions, properties of mesons are encoded in the quark-antiquark scattering matrices. The corresponding bound-state equation for mesons is the homogenous Bethe-Salpeter equation, which reads
\begin{equation}\label{HBSE}
\Gamma(k;P)=\int_q^{\Lambda}S_f(q_+)\Gamma(q;P)S_g(q_-)K(k,q;P)\,,
\end{equation}
where $P^2=-M^2$ with the meson mass $M$; $q_+=q+\alpha P$ and $q_-=q-(1-\alpha) P$ with the momentum fraction $\alpha\in[0,1]$; $f$ and $g$ denote the quark and antiquark flavors; $\Gamma(k,P)$ is the Bethe-Salpeter amplitude (BSA); $K(k,q;P)$ is the quark-antiquark scattering kernel. The dressed quark propagators are fed with the solution of the gap equation.

According to the $J^P$ quantum numbers, the BSA of the pseudoscalar ($J^P=0^-$) and vector ($J^P=1^-$) mesons can be expanded with a set of basis~\cite{Krassnigg2009PRD}:
\begin{eqnarray}
\Gamma_{0^-}(k;P)&=&\sum_{i=1}^{4}\tau_{0^-}^i(k;P) \mathcal{F}_i(k^2,z_k;P^2)\label{BSA_PS},\\
\Gamma_{1^-}(k;P)&=&\sum_{i=1}^{8}\tau_{1^-}^i(k;P) \mathcal{F}_i(k^2,z_k;P^2)\label{BSA_VC}.
\end{eqnarray}
The pseudoscalar basis is written as \cite{Chen2017PRD}
\begin{align}\label{Basis_PS}
 \tau_{0^-}^1&=i\gamma_5,        & \tau_{0^-}^2&=\gamma_5 \slashed{P}, \notag\\
 \tau_{0^-}^3&=\gamma_5\slashed{k}(k\cdot P),      &\tau_{0^-}^4&=i\gamma_5\sigma^{P,k}.
\end{align}
The vector basis reads~\cite{Chen2017PRD}
\begin{align}\label{Basis_VC}
\tau_{1^-}^1&=i\gamma_{\mu}^T,     &\tau_{1^-}^2&=ik_{\mu}^T\slashed{k},  \notag   \\
 \tau_{1^-}^3&=ik_{\mu}^T\slashed{P}(k\cdot P), &\tau_{1^-}^4&=\gamma_5\epsilon_{\mu\nu\alpha\beta}^T\gamma_{\nu}k_{\alpha}P_{\beta} ,     \notag        \\
 \tau_{1^-}^5&=k_{\mu}^T,      &\tau_{1^-}^6 &=\sigma_{\mu\nu}^T k_{\nu}(k\cdot P),   \\
 \tau_{1^-}^7&=\sigma_{\mu\nu}^T P_{\nu}      &\tau_{1^-}^8 &=k_{\mu}^T\sigma_{\alpha\beta}^T k_{\alpha}P_{\beta}\,, \notag
\end{align}
where $l_{\mu}^T = P_{\mu\nu}l_{\nu}$ with $P_{\mu\nu}=\delta_{\mu\nu}-\frac{P_{\mu}P_{\nu}}{P^2}$ as the transverse projector on the meson momentum $P$.

The $P$ and $C$ transformation for the BSA are defined as follows,
\begin{align}
\Gamma(k;P)&\stackrel{P}{\longrightarrow}\hat{P}\Gamma(\tilde{k};\tilde{P})\hat{P}^{-1} \label{PTrans},\\
\Gamma(k;P)&\stackrel{C}{\longrightarrow}\bar{\Gamma}(k;P)=\hat{C}\Gamma^t(-k;P)\hat{C}^{-1}\label{CTrans},
\end{align}
with $\tilde{k}=(k_4,-\vec{k})$, $\hat{P}=\gamma_4$, and $\hat{C}=\gamma_2\gamma_4$. The basis are chosen to have specific $P$- and $C$-parity. The coefficient functions $\mathcal{F}_i(k^2,z_k;P^2)$ can be expanded in terms of the Chebyshev polynomials
\begin{equation}\label{ChebyshevExpansion}
\mathcal{F}^{JP}(k^2,k\cdot P)=\sum_{j=0}a_j(k^2)U_j(k\cdot P/\sqrt{k^2P^2}).
\end{equation}

To normalize the BSA, we apply the Nakanish normalization condition~\cite{Nakanishi1965PRBI,Nakanishi1965PRBII}
\begin{equation}\label{Nakanishi Normalization}
\left(\frac{\partial \ln(\lambda)}{\partial P^2}\right)^{-1} = tr\int_k^{\Lambda}\bar{\Gamma}(k;-P)S(k_+)\Gamma(k;P)S(k_-),
\end{equation}
where $\bar{\Gamma}(k;-P)$ is the charge conjugation of $\Gamma(k;-P)$ as defined in Eq.~\eqref{CTrans}. With the normalized BSAs, the leptonic decay constants of pseudoscalar and vector mesons are defined as
\begin{eqnarray}
f_{0^-}P_{\mu} &=&\frac{Z_2}{\sqrt{2}} tr\int i\gamma_5\gamma_{\mu}S_f(k_+)\Gamma_{0^-}(k,P)S_g(k_-)\,,\label{fDecayPS}\\
f_{1^-} M  &=&\frac{Z_2}{3\sqrt{2}} tr\int \gamma_{\mu}S_f(k_+)\Gamma_{1^-}^{\mu}(k,P)S_g(k_-) \,. \label{fDecayVC}
\end{eqnarray}

\subsection{The Scattering Kernel}\label{SecIIC}

Now the scattering kernel is the only unknown piece of the whole puzzle. To construct it, we recall the axial-vector Ward-Takahashi identity (AV-WTI)~\cite{Maris1998PLB,Maris1997PRC} derived from the QCD chiral transformation, which reads
\begin{align}\label{AV-WTI}
P_{\mu}\Gamma_{5\mu}^{fg}(k;P)+i(m_{f}&+m_{g})\Gamma_{5}^{fg}(k;P)\notag\\
                                      & =S_{f}^{-1}(k_{+})i\gamma_{5}+i\gamma_{5}S_{g}^{-1}(k_{-})\,,
\end{align}
where $\Gamma_{5\mu}^{fg}(k;P)$ and $\Gamma_{5}^{fg}(k;P)$ are the axial-vector and pseudoscalar vertex, respectively, with two quark flavors denoted by $f$ and $g$. Note that the identity is model-independent and scheme-independent.

In the chiral limit, the current mass term vanishes. Inserting the Lorentz structure of the pion BSA, i.e.,
\begin{align}
\Gamma_{\pi}(k;P)=&\gamma_5[ i E_\pi(k;P)+\slashed{P} F_\pi(k;P)\notag\\
            &+\slashed{k}(k\cdot P)G_\pi(k;P)+\sigma_{\mu\nu}k_{\mu}P_{\nu}H_\pi(k;P)],
\end{align}
into the AV-WTI in Eq.~\eqref{AV-WTI}, we can obtain
\begin{equation}\label{AV-WTI-1}
f_{\pi}E_{\pi}(k;0)=B(k^2)\,,
\end{equation}
where $B(k^{2})$ is the scalar function defined in Eq.~\eqref{quark_lorentz2}.
It is apparent that the two-body problem is related to the one-body one and reveals that the pion can exist if and only if the chiral symmetry is dynamically broken.

The kernel with the one-gluon exchange form, which is called the ``ladder" approximation~\cite{Williams1994PPNP,Munczek1995PRD,Bender1996PLB}, can be expressed as ($l=k-q$)
\begin{equation}\label{Kernel}
K_{tu}^{rs}(q,k;P) = -\mathcal{G}(l^2) l^2 D_{\mu\nu}^{\text{free}}(l)\left(\gamma_{\mu}\frac{\lambda^a}{2}\right)_{tr}\left(\gamma_{\nu}\frac{\lambda^a}{2}\right)_{su}\,,
\end{equation}
where $t,u,r,s$ denote the Dirac and color indices. With the RL approximation, i.e., Eqs.~\eqref{InteractionModel} and \eqref{Kernel}, the solved quark propagator and vertices satisfy the identity in Eq.~\eqref{AV-WTI}. In other words, the RL approximation is a symmetry-preserving truncation scheme.

In order to construct a kernel beyond the ladder approximation, we first insert the gap equation and the inhomogeneous BSE into the AV-WTI, and then obtain
\begin{align}\label{Kernel_b2}
&\int_q K(q,k;P)[S_f(q_+)\gamma_5 + \gamma_5S_g(q_-)] \notag\\
                   & = -\int_q \gamma_{\mu} [D^f_{\mu\nu}(l)S_f(q_+)\gamma_5 + D^g_{\mu\nu}(l)\gamma_5S_g(q_-)]\gamma_{\nu}\,,
\end{align}
where the rainbow approximation has been used and all indices are suppressed for simplicity. Since the quark propagators with different flavors may have different effective interaction strengths, we use $D^{f}_{\mu\nu}$ and $D^{g}_{\mu\nu}$ for clarity.

\begin{figure}
 \centering
 \includegraphics[scale=0.5]{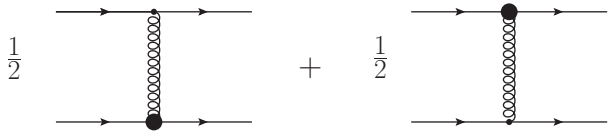}
 \caption{The Feynmann diagram for the kernel in Eq.~\eqref{KernelModify}. The black blobs denotes the elements in Eq.~\eqref{KernelModifyElem}.}
 \label{fig: KernelModifyFeyn}
 \end{figure}

Inspired by Eq.~\eqref{Kernel_b2}, we propose the BSE kernel as (the color matrices are suppressed)
\begin{align}\label{KernelModify}
K=-\frac{1}{2}\gamma_\mu\otimes\Sigma\cdot D_{\mu\nu}^{\Sigma}\cdot\gamma_\nu -\frac{1}{2}\gamma_\nu\cdot D_{\mu\nu}^{\Sigma}\cdot\Sigma\otimes\gamma_\mu,
\end{align}
with
\begin{align}\label{KernelModifyElem}
D_{\mu\nu}^{\Sigma}&=[D_{\mu\nu}^f(l) S_f(q_+) + D_{\mu\nu}^g(l) S_g(q_-)]\gamma_5, \notag\\
\Sigma &=\gamma_5[S_f(q_+) + S_g(q_-)]^{-1}.
\end{align}
The kernel can be illustrated by the Feynmann diagram as shown in Fig.~\ref{fig: KernelModifyFeyn}. If the two quarks have the same flavor, i.e., $D^{f} = D^{g}$, then the kernel degenerates into Eq.~\eqref{Kernel} and thus can be considered as a parameter-free generalization of the ladder approximation.

Straightforward analysis suggests that the kernel actually expresses an average of the effective interactions of the two flavors. For example, in the ultraviolet limit, i.e., $q \to \infty$, we obtain
\begin{equation}
K \rightarrow -\gamma\otimes\gamma\left(\frac{D^f+D^g}{2}\right),
\end{equation}
which is obviously a simple average of $D^{f}$ and $D^{g}$.
As another example, in the infrared limit, i.e., $q \to 0$, we have the kernel for $P \sim 0$ as
\begin{equation}
K \rightarrow -\gamma\otimes\gamma\left(\frac{D^f\sigma^f_{s}(0)+D^g\sigma^g_{s}(0)}
{\sigma^f_{s}(0)+ \sigma^g_{s}(0)}\right),
\end{equation}
which is also an average of $D^{f}$ and $D^{g}$ but weighted by the corresponding quark dressing functions.

%
\section{Results and Discussions}\label{SecIII}

\subsection{Light Quark Mesons}
Following Ref. \cite{Qin2018PRD}, for the mesons consisted of light quarks $u,d,s$, we take the interaction parameters as
\begin{equation}
\omega_q=0.5~\text{GeV}, \quad \varsigma_q=0.8~\text{GeV}.
\end{equation}
In the isospin symmetric limit: $m_{u} = m_{d}$, fitting with the underlined data in Table~\ref{table:mesonmassPS}, we take
\begin{equation}
m_{u/d}^{\zeta_{19}}=3.3~\text{MeV},\quad m_s^{\zeta_{19}}=74.6~\text{MeV},
\end{equation}
with which the masses and leptonic decay constants of light mesons are computed and listed in Tables~\ref{table:mesonmassPS} and ~\ref{table:mesonmassVC}. The renormalization-group-invariant masses are $\hat{m}_{u/d}=6.3$~MeV, $\hat{m}_s=146$~MeV. The one-loop-evolved masses at $2$~GeV are $m_{u/d}^{2\text{GeV}}=4.8~\text{MeV}$, $ m_s^{2\text{GeV}}=110~\text{MeV}$. The Euclidean constant quark masses are obtained
\begin{equation}
M_{u/d}^{E} = 0.41~\text{GeV}, \quad  M_s^{E} = 0.57~\text{GeV}\,,
\end{equation}
defined as $M_{q}^{E} =\{k|M_q(k^2)=k\}$, where $M_q(k^2)$ is the quark mass function.

\subsection{Heavy Quark Mesons}
For the heavy quark mesons, we follow the guide aforementioned and take the interaction parameters as
\begin{equation}
\omega_Q=0.8~\text{GeV}, \quad \varsigma_Q=0.6~\text{GeV}\,,
\end{equation}
which shape a much weaker effective interaction compared with that for the light quark mesons.

With the heavy quark current masses (fitted with the underlined data in Table~\ref{table:mesonmassPS})
\begin{equation}
m_{c}^{\zeta_{19}}=0.82~\text{GeV},\quad m_b^{\zeta_{19}}=3.59~\text{GeV},
\end{equation}
we obtain the masses and leptonic decay constants of of heavy quark mesons which are listed in Tables~\ref{table:mesonmassPS} and ~\ref{table:mesonmassVC}. The renormalization-group-invariant masses are $\hat{m}_{c}=1.61$~GeV, $\hat{m}_{b}=7.16$~GeV. The one-loop-evolved masses at $2$~GeV are $m_{c}^{2\text{GeV}}=1.22~\text{GeV}$, $m_b^{2\text{GeV}}=5.41~\text{GeV}$. The Euclidean constant quark masses are
\begin{equation}
M_{c}^{E} = 1.32~\text{GeV}, \quad M_b^{E} = 4.22~\text{GeV}.
\end{equation}

\subsection{Heavy-light Mesons}\label{SecIIIB}

With the parameters fixed in previous subsections, we can directly compute the properties of heavy-light mesons since the kernel beyond the ladder approximation, i.e., Eq. \eqref{KernelModify}, does not introduce any new parameters. In this work, we mainly focus on the mass spectrum and the leptonic decay constants.

\subsubsection{Mass Spectra}

\begin{figure}
 \centering
 \includegraphics[scale=0.33]{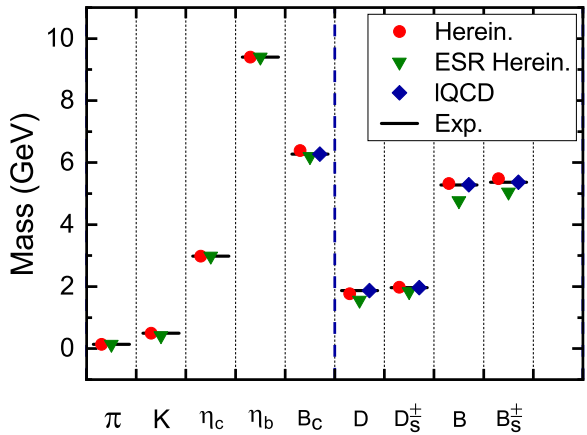}
 \caption{Masses of pseudoscalar mesons calculated herein compared with those from the equal spacing rule herein (ESR herein), the lQCD and the experiments. The vertical blue dashed line is used to separate the light-light and heavy-heavy mesons (in the left region) with the heavy-light mesons (in the right region). The detailed numbers are list in Table~\ref{table:mesonmassPS}.}
 \label{fig:mesonmassPS}
 \end{figure}
 \begin{figure}
 \centering
 \includegraphics[scale=0.35]{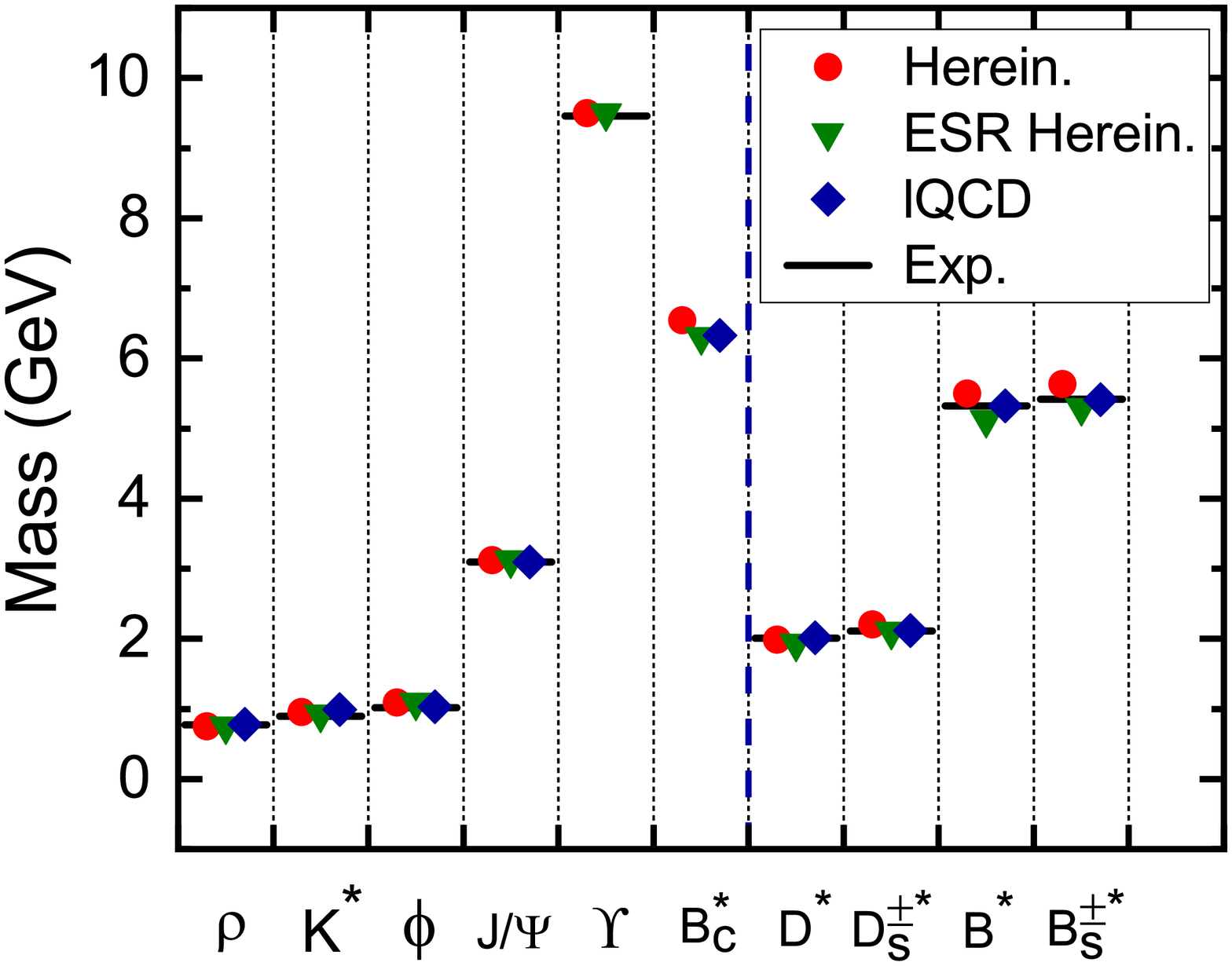}
 \caption{Masses of vector mesons calculated herein compared with those from the ESR herein, the lQCD and the experiments. The vertical blue dashed line is used to separate the light-light and heavy-heavy mesons (in the left region) with the heavy-light mesons (in the right region). The detailed numbers are list in Table~\ref{table:mesonmassVC}.}
 \label{fig:mesonmassVC}
 \end{figure}

\begin{table}
 \centering
\caption{Masses of pseudoscalar mesons calculated herein compared with those from the lQCD and experimental data. The ESR results are also presented for comparison. The lQCD results are taken from: $M_{B_c}$ - Ref.~\cite{Mathur2018PRL}; $M_D$ and $M_{D_s^{\pm}}$ - Ref.~\cite{Cichy2016PRD}; $M_B$ and $M_{B_s^{\pm}}$ - Ref.~\cite{Dowdall2012PRD}. The experimental data are from Ref.~\cite{PDG2018}. The underlined data are the experimental value used to fit the interaction parameters.}
 \label{table:mesonmassPS}
 \setlength{\tabcolsep}{2.5mm}{\begin{tabular}{c|c|c|c|c}
 \hline
 Meson        &Herein                &ESR Herein    &lQCD           &Exp.           \\ \hline
 $\pi$        &$\underline{0.138}$   &$0.138$       &$/$            &$0.138(1)$     \\
 $K$          &$\underline{0.495}$   &$0.416$       &$/$            &$0.495(1)$     \\
 $\eta_c$     &$\underline{2.984}$   &$2.984$       &$/$            &$2.984(1)$     \\
 $\eta_b$     &$\underline{9.399}$   &$9.399$       &$/$            &$9.399(1)$     \\
 $B_c$        &$6.388$               &$6.192$       &$6.276(7)$     &$6.275(1)$     \\
 $D$          &$1.771$               &$1.561$       &$1.868(3)$     &$1.868(1)$     \\
 $D_s^{\pm}$  &$1.981$               &$1.839$       &$1.968(4)$     &$1.968(1)$     \\
 $B$          &$5.324$               &$4.769$       &$5.283(8)$     &$5.279(1)$     \\
 $B_s^{\pm}$  &$5.478$               &$5.047$       &$5.366(8)$     &$5.367(1)$     \\ \hline
 \end{tabular}}
 \end{table}

\begin{table}
 \centering
 \caption{Masses of vector mesons calculated herein compared with those from the ESR herein, the lQCD and the experiment. The lQCD results are taken from: $M_{\rho}$ - Ref.~\cite{Fu2016PRD}; $M_{K^*}$ - Ref.~\cite{Dudek2014PRL}; $M_{\phi}$ - Ref.~\cite{Donald2014PRD}; $M_{J/\Psi}$ - Ref.~\cite{Donald2012PRD}; $M_{B_c}$ - Ref.~\cite{Mathur2018PRL}; $M_{D^*}$, $M_{D_s^{\pm*}}$, $M_{B^*}$ and $M_{B_s^{\pm*}}$ - Ref.~\cite{Lubicz2017PRD}. The experimental data are from Ref.~\cite{PDG2018}.}
 \label{table:mesonmassVC}
 \setlength{\tabcolsep}{2.5mm}{\begin{tabular}{c|c|c|c|c}
 \hline
 Meson        &Herein     &ESR Herein      &lQCD           &Exp.         \\ \hline
 $\rho$      &$0.749$    &$0.749$         &$0.780(16)$    &$0.775(1)$   \\
 $K^*$       &$0.953$    &$0.919$         &$0.993(1)$     &$0.896(1)$   \\
 $\phi$      &$1.089$    &$1.089$         &$1.032(16)$    &$1.019(1)$   \\
 $J/\Psi$    &$3.122$    &$3.122$         &$3.098(3)$     &$3.097(1)$   \\
 $\Upsilon$  &$9.497$    &$9.497$         &$/$            &$9.460(1)$   \\
 $B_c^*$     &$6.542$    &$6.310$         &$6.331(7)$     &$/$          \\
 $D^*$       &$1.988$    &$1.936$         &$2.013(14)$    &$2.009(1)$   \\
 $D_s^{\pm*}$&$2.206$    &$2.106$         &$2.116(11)$    &$2.112(1)$   \\
 $B^*$       &$5.501$    &$5.123$         &$5.321(8)$     &$5.325(1)$   \\
 $B_s^{\pm*}$&$5.635$    &$5.293$         &$5.412(6)$     &$5.415(2)$   \\ \hline
 \end{tabular}}
 \end{table}

We first compute the masses of the pseudoscalar and vector mesons. The comparisons with experimental data and lattice-QCD results are plotted in Fig.~\ref{fig:mesonmassPS} and Fig.~\ref{fig:mesonmassVC}.
The concrete data are summarized in Table~\ref{table:mesonmassPS} and Table~\ref{table:mesonmassVC}.
Among them, $\pi$, $K$, $\eta_c$, $\eta_b$ are taken as the calibration to determine the parameters.
Except the heavy-light mesons, i.e., $D$, $D^{\pm}_s$, $B$ and $B^{\pm}_s$, the kernels used for others are actually the ladder approximation because the interactions of the two flavors are the same in Eq.~\eqref{KernelModify}.

Following Ref.~\cite{Qin2018PRD}, we also include the results using the equal spacing rule (ESR). The key of the ESR is to define a constituent-quark passive-mass via
\begin{equation}
	M_X^f = \frac{1}{2}m_{X_{f\bar{f}}}\,,
\end{equation}
where $m_{X_{f\bar{f}}}$ denotes the mass of the $f$-flavor-symmetric meson in the channel $X$. The computed values (in GeV) are:
\begin{equation}
\label{EqMfPS}
{\rm PS:}\quad
\begin{array}{l|cccc}
f & u=d & s & c & b \\\hline
M_X^f & 0.07 & 0.35 & 1.49 & 4.70
\end{array}\,,
\end{equation}
and
\begin{equation}
\label{EqMfVC}
{\rm VC:}\quad
\begin{array}{l|cccc}
f & u=d & s & c & b \\\hline
M_X^f & 0.37 & 0.54 & 1.56 & 4.75
\end{array}\,,
\end{equation}
where PS, VC stands for the pseudo-scalar, vector mesons, respectively.
Then, the masses of flavor-unsymmetric mesons can be obtained by a straightforward interpolation as
\begin{eqnarray}
	m_{X_{f\bar{g}}} = M_X^f + M_X^g\,.
\end{eqnarray}

It is found that the masses of all pseudoscalar and vector mesons can be well reproduced with the BSE beyond the ladder approximation. The relative errors between the results herein and the experimental data are at the level of 5\%. For the $B_c^*$ meson, as the lack of the experimental result, we compare the result herein with the lattice-QCD and obtain the relative error only 4\%. On the other hand, the masses obtained with the ESR are comparable with the experimental data and the lattice-QCD, and the corresponding errors are just slightly larger than those with the BSE. Thus, the ESR can be an approximation for meson mass spectra.

As mentioned above, the kernel, i.e., Eq. \eqref{KernelModify}, actually expresses the averaging interaction of the light and heavy quarks. To demonstrate how the average works, we can define the effective arithmetic average as
\begin{equation}
K = -\gamma\otimes\gamma\left[\eta D^q+ (1-\eta) D^Q\right],
\end{equation}
where $\eta \in [0,1]$ is the weight parameter. For $\eta=1$, the kernel is the same as the one for flavor-symmetric light quarks. In this case, the interaction strength of the kernel is quite large and the binding between the light and heavy quarks are too strong, accordingly. The resulting meson mass is smaller than the experimental data. On the other hand, for $\eta=0$, the kernel is the same as the one for flavor-symmetric heavy quarks. In this case, the interaction strength of the kernel is rather small and the binding between the light and heavy quarks are too weak, accordingly. The resulting meson mass is heavier than the experimental data.

 \begin{figure}
 \centering
 \includegraphics[scale=0.4]{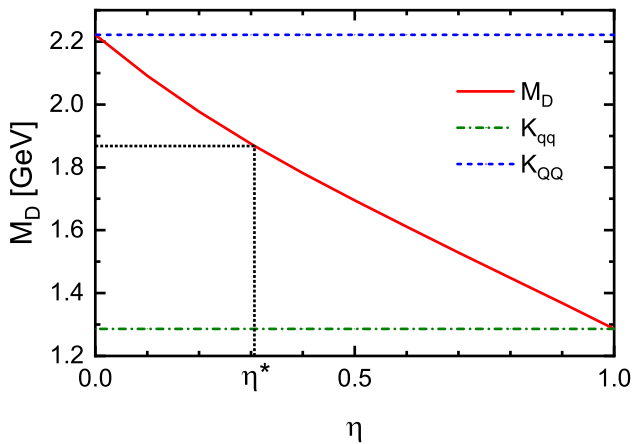}
 \caption{Masses of $D$ meson with the weight parameter of the arithmetic average of the kernel, where $\eta^*$ denotes the value corresponding to the experimental mass.}
 \label{fig:D_eta}
 \end{figure}

Now we can take $D$ meson as an example to analyze the mass with the running of the weight parameter. The result is shown in Fig.~\ref{fig:D_eta}. It is found that the weight parameter $\eta^*$ corresponding to the experimental mass indeed lies between the two limits, i.e., $0 < \eta^* < 1$. Thus, the kernel beyond the ladder approximation, i.e., Eq.~\eqref{KernelModify}, which produces the experimental mass, has the interaction strength equivalent to that weighted by $\eta^*$. However, Eq.~\eqref{KernelModify} is inspired by the AV-WTI and has no additional parameters. In other words, it expresses an implicit and automatic average. Moreover, the well-produced full mass spectra suggest that this kind of average is generally appropriate.

\subsubsection{Leptonic Decay Constants}

\begin{figure}
 \centering
 \includegraphics[scale=0.33]{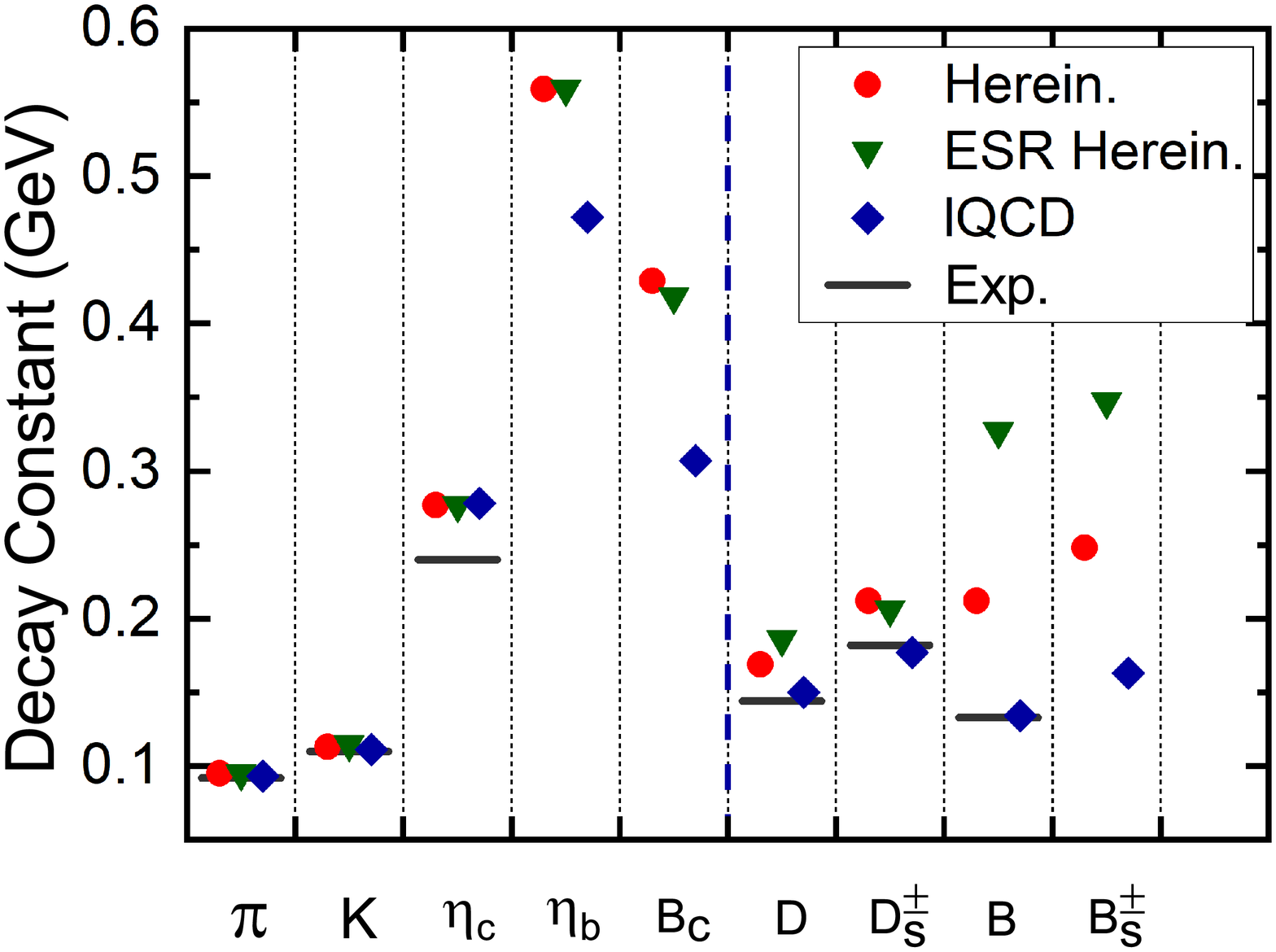}
 \caption{Decay constants of pseudoscalar mesons calculated herein compared with those from the ESR herein, the lQCD and the experiments. The vertical blue dashed line is used to separate the light-light and heavy-heavy mesons (in the left region) with the heavy-light mesons (in the right region). The detailed numbers are list in Table~\ref{table:fDecayPS}.}
 \label{fig:fDecayPS}
 \end{figure}

\begin{figure}
 \centering
 \includegraphics[scale=0.35]{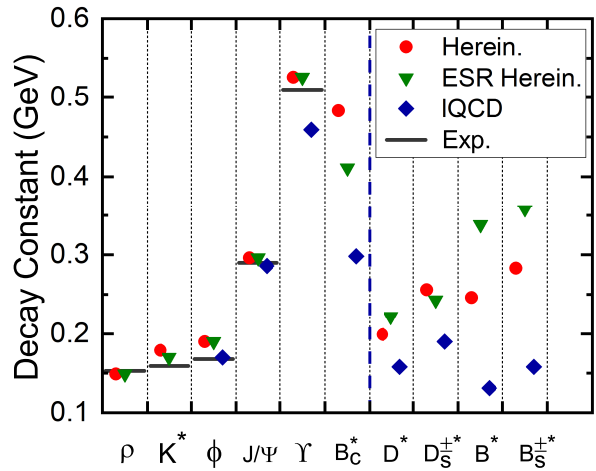}
 \caption{Decay constants of vector mesons calculated herein compared with those from the ESR herein, the lQCD and the experiments. The vertical blue dashed line is used to separate the light-light and heavy-heavy mesons (in the left region) with the heavy-light mesons (in the right region). The detailed numbers are listed in Table~\ref{table:fDecayVC}.}
 \label{fig:fDecayVC}
 \end{figure}

We then compute the leptonic decay constants of the pseudoscalar and the vector mesons. The obtained results are presented in Table~\ref{table:fDecayPS}, Table~\ref{table:fDecayVC}, respectively.
The comparisons with experimental data and lattice-QCD results are plotted in Fig.~\ref{fig:fDecayPS} and Fig.~\ref{fig:fDecayVC}.
Following Ref.~\cite{Qin2018PRD}, we also include the leptonic decay constants using the ESR,
which is defined as
\begin{eqnarray}
	f_{qQ} \approx f^{\rm interp}_{qQ} := \frac{1}{2} \left( f_{qq} + f_{QQ} \right) \,.
\end{eqnarray}

The first observation is that the leptonic decay constants are not as stable as the masses and the discrepancies from different sources become sizable for some mesons.
For the pseudoscalar mesons of light, strange, and charm flavors (i.e., $\pi$, $K$, $\eta_c$, $D$, and $D_s^\pm$), the theoretical methods produce similar values which are comparable with the experimental data. However the bottom-flavored pseudoscalar mesons (i.e., $\eta_b$, $B_c$, $B$, and $B_s^\pm$) lack of experimental data and have very different theoretical results. The values produced with the BSE and the ESR are larger than lattice-QCD ones.
Especially, for $B$ and $B_{s}^{\pm}$, we have $f_{\rm ESR} > f_{\rm BSE} > f_{\rm lQCD}$.
The reason why the ESR and the BSE produce large decay constants is that the interaction strength for the bottom sector is too strong since it is the same as that for the charm sector. This can be understood from the truth that $f_{\eta_b}$ obtained by the ESR and the BSE is significantly larger than that by lattice-QCD. If tuning $f_{\eta_b}$ down by decreasing the interaction strength for the bottom sector, the values of the BSE move close to the lattice-QCD results. However, the ESR still gives very large $f_{B}$ and $f_{B_s^\pm}$. This means that the ESR always fails for extremely flavor-unsymmetric mesons.

For vector mesons, the situation is analogous and the differences between the BSE/ESR and lattice-QCD results are generally similar to those for pseudoscalar mesons.
However, for the flavor-unsymmetric vector mesons, i.e., $D^*$ and $B^*$, the differences are significant. Note that $f_\Upsilon$ obtained by the BSE/ESR is larger than that by lattice-QCD just as $f_{\eta_b}$.
This is also a signature for that the interaction strength of the BSE/ESR is stronger than that of lattice QCD. If decreasing the interaction strength for the bottom sector as aforementioned, $f_\Upsilon$ decreases, and so does $f_{B^*}$. However, the differences between the BSE/ESR and lattice-QCD results cannot be removed completely and remain noticeable. The reason is that the kernel in Eq.~\eqref{KernelModify} only considers the AV-WTI which does not constrain the vector channel directly. Therefore, incorporating the axial-vector and vector WTIs to develop a universal BSE kernel for all channels could be the solution.

 \begin{table}
 \centering
 \caption{Decay constants of pseudoscalar mesons calculated herein compared with those from the ESR herein, the lQCD and the experiments. The lQCD results are taken from: $f_{\pi}$ and $f_K$ - Ref.~\cite{Follana2008PRL}; $f_{\eta_c}$, $f_{\eta_b}$ and $f_{B_c}$ - Ref.~\cite{McNeile2012PRD}; $D$, $D_s^{\pm}$, $B$, $B_s^{\pm}$ - Ref.~\cite{Bazavov2018PRD}. The experimental data are from Ref.~\cite{PDG2018,Edwards2001PRL}.
 }
 \label{table:fDecayPS}
 \setlength{\tabcolsep}{2.5mm}{\begin{tabular}{c|c|c|c|c}
 \hline
 Meson       &Herein        &ESR Herein       &lQCD          &Exp. \\ \hline
 $\pi$       &$0.095$       &$0.095$          &$0.093(1)$    &$0.092(1)$ \\
 $K$         &$0.113$       &$0.115$          &$0.111(1)$    &$0.110(1)$ \\
 $\eta_c$    &$0.277$       &$0.277$          &$0.278(2)$    &$0.237(52)$\\
 $\eta_b$    &$0.559$       &$0.559$          &$0.472(5)$    &$/$  \\
 $B_c$       &$0.429$       &$0.418$          &$0.307(10)$   &$/$  \\
 $D$         &$0.169$       &$0.186$          &$0.150(1)$    &$0.144(4)$ \\
 $D_s^{\pm}$ &$0.212$       &$0.206$          &$0.177(1)$    &$0.182(3)$  \\
 $B$         &$0.212$       &$0.327$          &$0.134(1)$    &$0.133(18)$  \\
 $B_s^{\pm}$ &$0.248$       &$0.347$          &$0.163(1)$    &$/$  \\ \hline
 \end{tabular}}
 \end{table}

 \begin{table}
 \centering
 \caption{Decay constants of vector mesons calculated herein compared with those from the ESR herein,
 the lQCD and the experiments. The lQCD results are taken from: $f_{\phi}$ - Ref.~\cite{Donald2014PRD}; $f_{J/\Psi}$ - Ref.~\cite{Donald2012PRD}; $f_{\Upsilon}$ - Ref.~\cite{Colquhoun2015PRD91.074514}; $f_{B_c^*}$ - Ref.~\cite{Colquhoun2015PRD91.114509}; $f_{D^*}$, $f_{D_s^{\pm*}}$, $f_{B^*}$, $f_{B_s^{\pm*}}$ - Ref.~\cite{Lubicz2017PRD}. The experimental data are from Ref.~\cite{PDG2018,Maris1999PRC}.}
 \label{table:fDecayVC}
 \setlength{\tabcolsep}{2.5mm}{\begin{tabular}{c|c|c|c|c}
 \hline
 Meson        &Herein         &ESR Herein          &lQCD              &Exp.   \\ \hline
 $\rho$       &$0.149$        &$0.149$             &$/$               &$0.153(1)$ \\
 $K^*$        &$0.179$        &$0.170$             &$/$               &$0.159(1)$ \\
 $\phi$       &$0.190$        &$0.190$             &$0.170(13)$       &$0.168(1)$ \\
 $J/\Psi$     &$0.296$        &$0.296$             &$0.286(4)$        &$0.294(5)$ \\
 $\Upsilon$   &$0.526$        &$0.526$             &$0.459(22)$       &$0.505(4)$ \\
 $B_c^*$      &$0.483$        &$0.411$             &$0.298(9)$        &$/$    \\
 $D^*$        &$0.199$        &$0.222$             &$0.158(6)$        &$/$    \\
 $D_s^{\pm*}$ &$0.256$        &$0.243$             &$0.190(5)$        &$/$    \\
 $B^*$        &$0.246$        &$0.338$             &$0.131(5)$        &$/$    \\
 $B_s^{\pm*}$ &$0.283$        &$0.358$             &$0.158(4)$        &$/$    \\ \hline
 \end{tabular}}
 \end{table}


 \section{Summary}\label{SecIV}
%
In summary, the heavy-light mesons in pseudoscalar and vector channels are investigated within the Dyson-Schwinger equation framework. The rainbow-ladder truncation is modified under the consideration of the symmetries of QCD. It combines the effects of the light and heavy sectors. Moreover, the modified kernel involves no new parameter and can degenerate into the RL truncation for flavor-symmetric heavy-heavy or light-light mesons.

With the kernel beyond the ladder approximation, the calculated masses of the heavy-light mesons in pseudoscalar and vector channels agree with the experimental data at the level of $5\%$. The masses obtained with the ESR are comparable with the experimental data and the lattice-QCD results while the corresponding errors are slightly larger than those with the BSE. From our analysis, the modified kernel expresses an implicit and automatic average with no additional averaging parameter. These facts shows that the modified kernel is appropriate to describe the mass spectrum of the ground states mesons.

For the leptonic decay constants, our calculated results are not as stable as those for the masses. For the pseudoscalar mesons of light, strange and charm flavors, our theoretical methods produce similar values that are comparable with the experimental data. While for the bottom-flavored mesons, both the ESR and the BSE produce larger values than the lattice-QCD. The reason is that the interaction strength for the bottom sector is too strong since it is the same as that for the charm sector. For vector mesons, the differences between the BSE/ESR and lattice-QCD results are generally analogous but larger than those for pseudoscalar mesons. Besides the interaction strength, the missing constraint from the vector Ward-Takahashi identity is a factor to explain the larger differences.

From the discussion above, we can infer that the kernel beyond the ladder approximation is a promising tool to depict the quark-antiquark interactions. The mass spectra and the leptonic decay constants of the light-light, heavy-heavy and heavy-light mesons in the pseudoscalar and vector channels can be well described, systematically. The modified kernel can be further improved by incorporating the axial-vector and vector Ward-Takahashi identities together and used to study meson and baryon properties. The related work is under progress.\\

\begin{acknowledgments}
The authors are grateful for the fruitful discussion with C.D. Roberts. The work was supported by the National Natural Science Foundation of China under Contracts No. 11435001, No. 11775041, No.11847301, No. 11805024, and the National Key Basic Research Program of China under Contract No. 2015CB856900.
This work was also supported in part by the Fundamental Research Funds for the Central Universities under Grant No. 2019CDJDWL0005.
\end{acknowledgments}


\bibliographystyle{unsrt}

\end{document}